\documentclass{article}
\usepackage{graphicx}
\begin{document}
\title{Einstein-Cartan cosmology and the high-redshift Universe}
\author{Davor Palle \\
ul. Ljudevita Gaja 35, 10000 Zagreb, Croatia \\
email: davor.palle@gmail.com}
\maketitle
\begin{abstract}
{The Hubble tension, known as a discrepancy between the local measurements
vs. the CMB, SNe and galaxy clustering fits of the Hubble constant, the first
measurement of the 21-centimeter high-redshift signal by EDGES, the
high-redshift galaxy halo number densities and the measurements of the ionizing 
photon mean free path represent a great challenge
for the concordance cosmology.
We show that the nonsingular Einstein-Cartan cosmological model with 
the simple parametrization of torsion of spacetime (angular momentum of
the Universe) can substantially improve agreement with data.
Light Majorana neutrinos are dominant source of the spin of matter coupled to
torsion, while the heavy Majorana neutrinos represent cold dark matter
particles fulfilling the Griest-Kamionkowski unitarity bound.}
\end{abstract}                                        
  
\section{Introduction and motivation}
The General Relativity (GR), the cosmological constant and the cold dark matter
(CDM) are the fundamental building blocks of the $\Lambda$CDM concordance cosmological model adopted to reproduce satisfactorily a large number of cosmological phenomena \cite{Peebles,Kolb,Paddy,Peacock}. However, abundant new
measurements of high-redshift quasars and galaxies, 21-centimeter absorption
signal, precise local Hubble constant measurements against those inferred from
the CMB, lensing, SNe or galaxy clustering and a notorious lack of the large scale power of CMB spectrum, dictate the inevitable need to substantially
improve or even completely replace the $\Lambda$CDM model.

In this paper we explicate and illuminate the possibility how to comprehend
the appearance of dark matter and dark energy starting from the
nonsingular and causal theories in the local Minkowski frame and in the 
global curved space-time. The subsequent chapters are devoted to a description
of the particle content of the Universe, demonstrating also the difference from 
the Standard Model (SM) of particle physics; to introduction of the Einstein-Cartan
cosmology compared to $\Lambda$CDM model and finally to explore the observables of
the high-redshift Universe.

\section{Particle content of the Universe}
Intending to remove the ultraviolet-UV (zero-distance) singularity and 
the SU(2) global anomaly, a new theory of the local relativistic quantum fields
is formulated in \cite{Palle1}. We show in \cite{Palle1} the intricate
relations between gauge, discrete and conformal symmetries in Minkowski
space-time and how to acquire electromagnetic, weak and strong interactions
embedded into the conformal SU(3) gauge symmetry.

The spectrum of the theory can be revealed within the study of 
Dyson-Schwinger equations with the UV cut-off defind by the mass
of weak gauge W and Z bosons 
$\Lambda = \frac{\pi}{\sqrt{6}}\frac{2}{g} M_{W}$, 
 $e = g \sin \Theta_{W}$, $\cos \Theta_{W} = \frac{M_{W}}{M_{Z}}$.
Instead of the SM massless Dirac neutrinos, the theory of ref. \cite{Palle1}
contains light and heavy Majorana neutrinos. The weak boson mixing angle
and quark (lepton) mixing angles must fulfill the following relation
in order to preserve a cancelation of the fermion and boson SU(2) global
anomalies \cite{Palle1}:

\begin{eqnarray}
\Theta_{W}=2 (\Theta_{12}^{D}+\Theta_{23}^{D}+\Theta_{31}^{D}).
\end{eqnarray}

This relation
must be valid even for Majorana neutrinos. Namely, in the case of the inverted
mass hierarchy, for example if $m_{\nu,1}^{M} > m_{\nu,2}^{M}$ for 
light Majorana neutrinos,
$m_{N,1}^{M} < m_{N,2}^{M}$ for heavy Majorana neutrinos
and $m_{\nu,1}^{D} < m_{\nu,2}^{D}$ for Dirac neutrinos, the see-saw mechanism and the Euler
matrix imply $\Theta_{12}^{D}=-\Theta_{12}^{M}$:

\begin{eqnarray*}
\left( \begin{array}{cc}
\cos \Theta_{12} & \sin \Theta_{12} \\
-\sin \Theta_{12}  & \cos \Theta_{12} \end{array}
\right)
\left( \begin{array}{c}
u_{1} \\
u_{2} \end{array}
\right), 
\end{eqnarray*}

is equivalent to:

\begin{eqnarray*}
\left( \begin{array}{cc}
\cos \Theta_{12} & -\sin \Theta_{12} \\
\sin \Theta_{12}  & \cos \Theta_{12} \end{array}
\right)
\left( \begin{array}{c}
u_{2} \\
u_{1} \end{array}
\right).
\end{eqnarray*}

\newpage

Before checking the relation between boson and fermion mixing angles,
one has to verify the unitarity of the quark and lepton mixing matrices.

Very strong coupling of heavy Majorana neutrinos and Nambu-Goldstone
scalars ensures very large masses \cite{Palle1}. Light and heavy 
Majorana neutrinos are cosmologically stable particles, i.e. 
$\tau_{N_{i}},\tau_{\nu_{j}}  > \tau_{U}$ \cite{Palle2}.
Furthermore, heavy Majorana neutrinos are perfect candidates for 
the CDM particles \cite{Griest} that simultaneously solve
the leptogenesis \cite{Japovi} and baryogenesis problems since 
$\Delta L=\Delta B$ \cite{Kolb} in the (denoted as BY) theory 
of ref. \cite{Palle1}. CP violation in lepton and baryon processes
is mandatory \cite{Sakharov}.

The particle content of the BY theory has all cosmological
ingredients such as CDM, small amount of hot dark matter (light
neutrinos) and lepton (baryon) number violation, therefore just
everything what the SM does not contain.

The UV cut-off $\Lambda = 326 GeV = 0.6\times 10^{-16} cm$
as a dimensionful parameter of the noncontractible space
breaks the gauge, conformal and discrete symmetries, and
it causes profound phenomenological consequences in particle 
physics. Flavour anomalies in suppressed B meson decays
\cite{Palle3,Palle4} or few  QCD anomalies \cite{Palle5,Palle6}
are a matter of intense experimental investigations.

The proclaimed discovery of the 125 GeV Higgs scalar of the SM
cannot be considered as a final assertion. Namely, 125 GeV resonance
can be interpreted as a scalar meson by coupling scalar toponium and 
gluonium bound states \cite{Cea,Palle7}. The additional 750 GeV
scalar resonance can be detected only through its small gluonium part
since it is above toponium threshold (346 GeV).
Namely, it is very well known that
t-quark decays faster than the occurence of the hadronization process. 
Therefore, two reasons for the very small number of 750 GeV resonance events:
six times heavier than 125 GeV scalar and the visibility of only smaller gluonium
part, results in only $\simeq3 \sigma$ significance with Run1 and Run2 data.
Forthcoming Run3 and high-luminosity LHC should resolve the nature of
both 125 GeV and 750 GeV resonances.

Similar type of hadron resonances
can be expected from the mixture of bottomonium and gluonium. It is
advantageous that numerical simulations with lattice QCD and b-quarks
are feasible. Assuming the strong coupling of heavy quarkonia and
gluonium (i.e. large annihilation matrix element \cite{Palle7})
and $A(t\bar{t},gg)/m(t\bar{t}) \simeq A(b\bar{b},gg)/m(b\bar{b})$,
the scalar bottomonium-gluonium $\simeq$30 GeV heavy resonance should appear
\cite{Heister}.

\section{Introducing the Einstein-Cartan cosmology}
Initial conditions of the singular $\Lambda$CDM cosmology are established 
with the introduction of the inflaton scalar \cite{Kolb,Peacock}. 
The parameters of inflation are fine tuned to solve various problems 
of the expanding GR cosmology. The total density normalized
to the critical density is predicted to be $\Omega_{tot}=1$, but the mass
density and cosmological constant are obtained as a fit to data, not as 
a firm prediction of the inflation and $\Lambda$CDM cosmology.

The Einstein-Cartan theory of gravity (cosmology) presents the generalization
of the GR by including rotational degrees of freedom \cite{Sciama}.
The first important cosmological result for Einstein-Cartan(EC) cosmology
was proved by A. Trautman \cite{Trautman} demonstrating how to avoid
a singularity by a torsion.

To envisage the evolution of the Universe from some $R_{min}$ to $R=\infty$,
we recall the reader to, for example, integral Dyson-Schwinger equation in
Minkowski space-time for some fermion. Any solution of this integral 
equation in the space-like domain ($ 0 \leq -q^{2} \leq \Lambda^{2}$)
at some particular space-time point is defined by a behaviour in the whole 
space-like domain. This is a bootstrap character of Dyson-Schwinger integral
equation. A continuation to the time-like domain to gain the spectrum
of the fermion requires solutions in both space-like and time-like
domains. We suspect that some kind of bootstrap exist within EC cosmology.

Verifying that the EC cosmology has $R_{min}\simeq {\cal O} (10^{-16}cm)$
\cite{Palle8}, we show that quantum fluctuations at $R_{min}$ generate 
primordial density fluctuations \cite{Palle9,Palle10}.
During the era of the CDM (heavy Majorana neutrinos) decoupling,
the abundant number of right-handed helicity(chirality) light neutrinos are
created \cite{Palle11}. Important to restate that heavy Majorana 
neutrinos are responsible for lepton and baryon number violations \cite{Japovi}.
All particles must fulfill EC equations of motion \cite{Palle8,Palle10}:

\begin{eqnarray}
&&\frac{d^{2}x^{\mu}}{ds^{2}} + \Gamma^{\mu}_{(\nu\kappa)}\frac{dx^{\nu}}{ds}\frac{dx^{\kappa}}{ds}=0, \\ 
&& (\mu\nu)=\frac{1}{2}(\mu\nu+\nu\mu),\ \ 
\Gamma^{\mu}_{(\nu\kappa)}=\left\{ \begin{array}{c} \mu \\ \nu\kappa \end{array}\right\}+ Q^{\ \ \mu}_{\nu\kappa .}+Q^{\ \ \mu}_{\kappa\nu .},  \nonumber \\
&&\Gamma^{\mu}_{\nu\kappa}=\left\{ \begin{array}{c} \mu \\ \nu\kappa \end{array}\right\}+Q^{\     \ \mu}_{\nu\kappa .}
+Q^{\ \ \mu}_{\kappa\nu .}+Q^{\mu}_{. \nu\kappa}, \nonumber \\
&& torsion\ tensor = Q^{\mu}_{. \nu\kappa} = \frac{1}{2}(\Gamma^{\mu}_{\nu\kappa}-\Gamma^{\mu}_{\kappa\nu})
= u^{\mu}Q_{\nu\kappa}, \nonumber \\
&& u^{\mu}= velocity\ fourvector,\ 
Q^{2} = \frac{1}{2}Q_{\mu\nu}Q^{\mu\nu}. \nonumber
\end{eqnarray}

Because of the right-handed chirality light neutrino number and spin density dominance at
early times of evolution
and EC equations of motion with torsion term, violation of isotropy 
inevitably occurs. Averaged spin (torsion) of matter (space-time) 
cannot be erased in the ongoing evolution. Small at early times,
after creation of larger and larger structures (stars, globular clusters,
galaxies, clusters, superclusters), torsion is now proportional to the angular momentum
of the Universe with right handed chirality \cite{Palle11}.

Let us denote the cosmic scale factor by $a(z) = 1/(1+z)= R(t)/R_{0}$.
The evolution from the present state to infinity is then for
$\ a \epsilon [1,\infty ),\ z \epsilon (-1,0]$. Acknowledging the 
state of the Universe in the cosmology in the vicinity of 
spacelike infinity \cite{Obukhov} and conformal technique of Penrose
\cite{Penrose} we come to the following conclusions \cite{Palle8,Palle10}:

\begin{eqnarray}
&&\lim_{R \rightarrow \infty} \frac{\rho_{m}}{\rho_{crit}} = 2,\ \ 
\lim_{R \rightarrow \infty} \frac{\rho_{Q}}{\rho_{crit}} = -1,\ \ 
\lim_{R \rightarrow \infty} \frac{\rho_{\Lambda}}{\rho_{crit}} = 0,  \\
&&\Omega_{cmb}={\cal O}(10^{-4}) \ll 1 \Rightarrow
\Omega_{m}=2,\ \ \Omega_{Q}=-1,\ \ \Omega_{\Lambda}=0. \nonumber
\end{eqnarray}

Precisely at this point one can observe the action of bootstrap: matter
density contributes to the curvature but the negative opposing contribution of
the angular momentum of matter through torsion-bootstrap action infers to  
equality of the total and critical densities.

Summarizing all results of the EC comology we conclude that there is no need
for the inclusion of the inflaton scalar (after all $R_{min}$ is at the weak
interaction scale not at the Planck scale) since we can generate primordial
perturbations and anisotropy through torsion. Moreover, we fix even matter density.

We need to model the evolution of torsion (angular momentum) of the Universe in
order to perform numerical evaluations in the next chapter.
The most natural method is to equate the ages of the Universe for $\Lambda$CDM
and EC cosmologies as our cosmic chronometers:

\begin{eqnarray*}
&&\Lambda CDM: \tau_{U} = \frac{9.7776}{h}\int^{1}_{10^{-3}}\frac{da}{a}
[\Omega_{m}a^{-3}+\Omega_{\Lambda}]^{-\frac{1}{2}}Gyr,  \\ 
&& EC: \tau_{U} = \frac{9.7776}{h}\int^{1}_{10^{-3}}\frac{da}{a}
[\Omega_{m}a^{-3}-\frac{1}{3}Q^{2}(a)]^{-\frac{1}{2}}Gyr,  \\
&&\tau_{U}(EC) \simeq  \tau_{U}(\Lambda CDM)  \\   &&   \Rightarrow 
\frac{1}{h_{1}}[\Omega_{m,1}a^{-3}-\frac{1}{3}Q^{2}(a)]^{-\frac{1}{2}}
\simeq \frac{1}{h_{2}}[\Omega_{m,2}a^{-3}+1-\Omega_{m,2}]^{-\frac{1}{2}} \\
&& \Rightarrow Q(a)=\sqrt{3}[(\frac{h_{2}}{h_{1}})^{2}(\Omega_{m,2}-1)
+(\Omega_{m,1}-\Omega_{m,2}(\frac{h_{2}}{h_{1}})^{2})a^{-3}]^{\frac{1}{2}} \\
&& \Rightarrow Q(a)=\sqrt{3}[-c_{1}+c_{2}a^{-3}]^{\frac{1}{2}}\simeq
\sqrt{3c_{2}}a^{-\frac{3}{2}}[1-\frac{c_{1}}{2c_{2}}a^{3}+...].
\end{eqnarray*}

We neglect very small terms with vorticity \cite{Palle10}. The resulting
expression for angular momentum coincides with the evolution of the
angular momentum of galaxies in the Zeldovich model \cite{Paddy}.
Our final parametrization of torsion has the form (two redshifts are introduced
to simulate a smooth rise of torsion):

\begin{eqnarray*}
Q(a)&=&\sqrt{3}(1 + \frac{a-a_{0}}{a_{0}-a_{1}})[1-c+c a^{-3}]^{\frac{1}{2}},\ 
for\ a_{1} \leq a \leq a_{0}, \\
a&=&\frac{1}{1+z},\ a_{0}=\frac{1}{1+z_{0}},\ a_{1}=\frac{1}{1+z_{1}}, \\
Q(a)&=&\sqrt{3}[1-c+c a^{-3}]^{\frac{1}{2}},\ for\ a_{0} \leq a \leq 1, \\
\Omega_{Q}&=& -1\ \Longleftrightarrow \ Q(a=1)=\sqrt{3},\ \ 
Q(a \leq a_{1}) = 0 .
\end{eqnarray*}

In Table 1 we summarize parameters of models ($\tau_{U}(EC)=13.92\ Gyr,\ \tau_{U}(\Lambda CDM)=13.83\ Gyr$).

\begin{table}
\caption{Model parameters.}
\vspace{3mm}
\hspace{15mm}
\begin{tabular}{| c || c | c | c | c | c | c |} \hline
model & $\Omega_{m}$ & $\Omega_{\Lambda}$ & h & c & $z_{0};z_{1}$ & $\Omega_{b}$ \\  
\hline \hline
$\Lambda$CDM & 0.307 & 0.693 & 0.677  & -  & -  & 0.045\\  \hline 
EC  &  2  & 0  &  0.74  &  1.85  &  4;6 &  0.045   \\  \hline
\end{tabular} 
\end{table}

The equation for the growth function $ D_{+}(a)$ within the EC cosmology can be
easily derived \cite{Palle10}:

\begin{eqnarray}
&&\Lambda CDM:\ \delta (a)= \frac{\sqrt{1+x^{3}}}{x^{3/2}}
\int^{x}_{10^{-3}x_{0}} \frac{dx x^{3/2}}{(1+x^{3})^{3/2}}, \nonumber  \\
&& x=x_{0} a,\ x_{0}=(\frac{\Omega_{\Lambda}}
{\Omega_{m}})^{1/3},  \nonumber \\
&&EC : \frac{d^{2}\delta}{d a^{2}}=(\dot{a})^{-2}[-(\ddot{a}+2 \frac{\dot{a}^{2}}{a})
\frac{d \delta}{d a}+\frac{3}{2}\frac{\dot{a}^{2}}{a^{2}}\delta], \\ 
&&\frac{\dot{a}^{2}}{a^{2}}=H_{0}^{2}[\Omega_{m}a^{-3}-\frac{1}{3}Q^{2}(a)], \nonumber \\ 
&& \frac{\ddot{a}}{a}=H_{0}^{2}[-\frac{1}{2}\Omega_{m}a^{-3}+\frac{2}{3}Q^{2}(a)], \nonumber \\
&& D_{+}(a) \equiv \frac{\delta (a)}{\delta (1)},\ H_{0}=100 h\ km s^{-1}Mpc^{-1}. \nonumber
\end{eqnarray}

Figure 1 describes 
the quotient of the growth functions for EC and $\Lambda$CDM models.

\begin{figure}[htb]
\centerline{
\includegraphics[width=15cm]{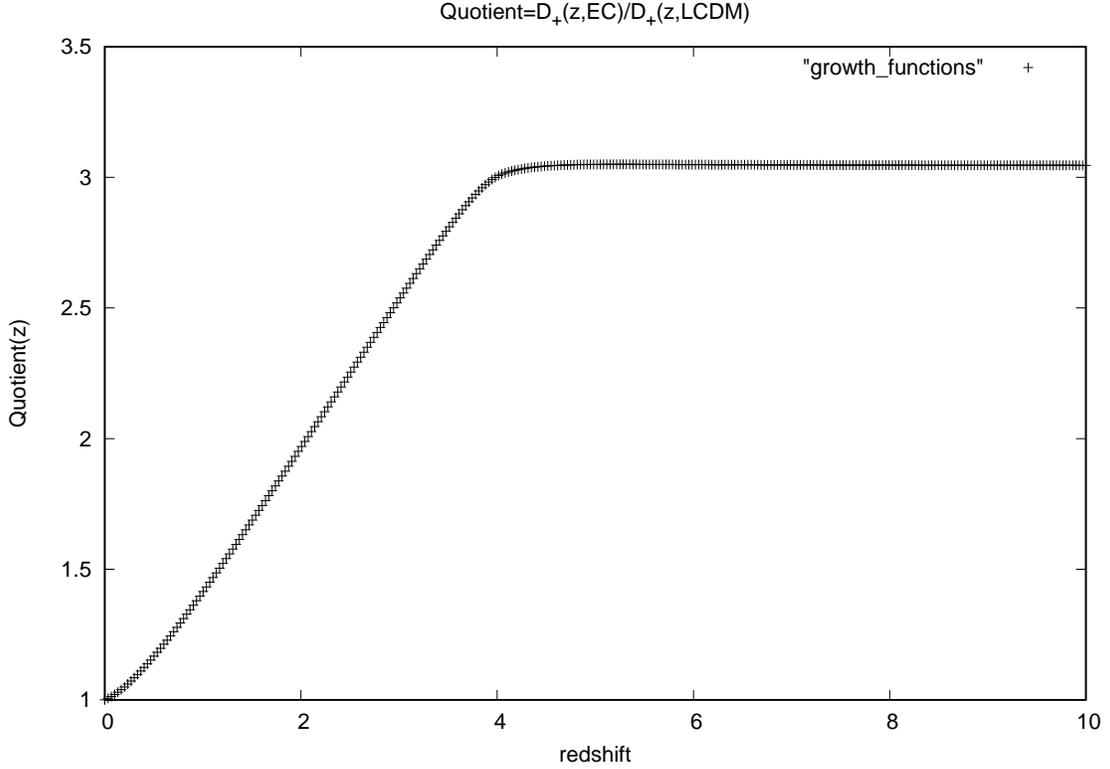}}
\caption{Quotient of growth functions $D_{+}(z,EC)/D_{+}(z,\Lambda CDM)$}
\end{figure}

\section{High-redshift Universe}
With a rapid growth of our knowledge based on new measurements and more
precise theoretical computations, cosmologists are faced with the first
large discrepancy of the $\Lambda$CDM model: accurate measurements
of the Hubble parameter in our "local" (very small redshift) Universe \cite{tension}
vs. fit of the Hubble parameter with CMB data (evolution from $z_{dec}
\simeq 1100$ to $z=0$), with galaxy clustering at various redshift, with gravitational
lenses, etc. This is a clear evidence that the $\Lambda$CDM model with cold dark matter, 
cosmologial constant and isotropic and homogeneous FRW geometry does not describe
the physical reality any more. We mentioned already the lack of power of the TT CMB
spectrum at large scales. This topic deserves a separate investigation, however it is
no more a surprise in light of the "Hubble tension" and other high-redshift
problems.

We choose an observable that is well measured and redshift dependent: the galaxy
halo number density. The evaluation requires our choice of primordial density
spectrum, processed spectrum imprinted in transfer functions,  
fitting functions and growth functions. The Harrison-Zeldovich-Peebles primordial
spectrum \cite{Peebles,Kolb,Paddy,Peacock}
with a small tilt \cite{WMAP,Palle12,Planck1} is applied:

\begin{eqnarray*}
P(k,z) &=& C P_{i}(k) T^{2}(k) D_{+}^{2}(z),\ C=normalization\ constant, \\
P_{i}(k) &=& k^{\alpha},\ \alpha=0.96,\ T(k)=transfer\ function. 
\end{eqnarray*}

Since both $\Lambda$CDM and EC models assume the adiabatic CDM perturbations,
we use the corresponding transfer function as a result of the integrated Boltzmann
equations with CDM, baryons and photons (small corrections of light neutrinos are
neglected)\cite{Peacock}:

\begin{eqnarray*}
T(k)&=&\frac{\ln (1+2.34 q)}{2.34 q}[1+3.89 q+(16.1 q)^{2} 
+(5.46 q)^{3}+(6.71 q)^{4}]^{-1/4}, \\
q&=&\frac{k}{(\Gamma h Mpc^{-1})},\ \Gamma=h \Omega_{m} exp[-\Omega_{b}
(1+\sqrt{2 h}/\Omega_{m})].
\end{eqnarray*}

The fitting functions are defined by probability calculations of high-density
regions \cite{Paddy} and we choose the Sheth-Tormen function because of its
universal applicability with respect to halo mass and redshift \cite{Sheth}:

\begin{eqnarray*}
f_{ST}(\sigma)&=&A\sqrt{\frac{2r}{\pi}}[1+(\frac{\sigma^{2}}{r\delta^{2}_{c}})^{p}]
\frac{\delta_{c}}{\sigma} exp[-\frac{r\delta^{2}_{c}}{2\sigma^{2}}], \\
A&=&0.3222,\ \delta_{c}=1.686,\ r=0.707,\ p=0.3, \\
&&\int^{+\infty}_{-\infty}f_{ST} (\sigma) d \ln \sigma^{-1} = 1.
\end{eqnarray*}

The growth functions are evaluated in the preceding section and we are now
equipped with all the necessary ingredients to evaluate galaxy halo number density
as a function of halo mass and redshift \cite{Peacock}:

\begin{eqnarray}
&&\frac{d n(M,z)}{d M} = f_{ST}(\sigma)\frac{\bar{\rho}_{m}(z=0)}{M}
\frac{d \ln \sigma^{-1}}{d M}, \\
&&\bar{\rho}_{m}(z)=\Omega_{m}(1+z)^{3}\rho_{crit}(z=0), 
 \ M=\frac{4\pi\bar{\rho}_{m}(z=0)}{3}R^{3}, \nonumber  \\
&&\sigma^{2}(R,z)=\frac{1}{2\pi^{2}}\int^{+\infty}_{0}dk\ k^{2}P(k,z)W^{2}(kR), \nonumber  \\
&&\rho_{crit}(z=0)=\frac{3 H_{0}^{2}}{8\pi G_{N}},\ \sigma=variance, \nonumber  \\
&&W(x)= window\ function=\frac{3}{x^{3}}(\sin x - x \cos x). \nonumber  
\end{eqnarray}

It is easy to verify that:

\begin{eqnarray*}
&&\frac{d \ln \sigma^{-1}}{d M}=-\frac{3}{2\pi^{2}M\sigma^{2}R^{4}}
\int^{+\infty}_{0}dk\ k^{-2}P(k,z) \\ &&\times (\sin kR-kR\cos kR)
[\sin kR(1-\frac{3}{(kR)^{2}})+3\frac{\cos kR}{kR}].
\end{eqnarray*}

We present results for $\Lambda$CDM and EC models normalizing the spectrum
by the standard method with a variance $\sigma_{8} = \sigma (R_{8}=h^{-1}8 Mpc,z=0)
= 0.823$.

The evaluations of the halo number density for halo masses from $10^{9}$ to $10^{12} M_{\odot}$
and for redshifts from $z=4$ to $z=8$ depicted in Table 2 and Figure 2 clearly show
large enhancements in EC cosmology with respect to $\Lambda$CDM model.

\begin{table}
\caption{Halo number density as a function of mass and redshift; 
upper number=$\frac{d n}{d M}(EC;Mpc^{-3}M_{\odot}^{-1})$, lower number=$\frac{d n}{d M}(EC)$
/$\frac{d n}{d M}(\Lambda CDM)$.}
\vspace{3mm}
\hspace{5mm}
\begin{tabular}{| c || c | c | c | c |} \hline
        &             &                   &             &           \\
z $\backslash$ M($M_{\odot}$) & $10^{9}$ & $10^{10}$& $10^{11}$& $10^{12}$ \\  
\hline \hline
  &$4.06\cdot 10^{-9}$&$5.59\cdot 10^{-11}$  & $8.09\cdot 10^{-13}$& $1.25\cdot 10^{-14}$  \\  
 4 & 4.26 & 6.38 & 13.18  & 55.61 \\  \hline
  &$4.78\cdot 10^{-9}$&$6.60\cdot 10^{-11}$  & $9.57\cdot 10^{-13}$& $1.47\cdot 10^{-14}$  \\
 6 & 7.79 & 16.66 & 66.29  & 1048.36 \\  \hline
  &$5.45\cdot 10^{-9}$&$7.53\cdot 10^{-11}$  & $1.09\cdot 10^{-12}$& $1.65\cdot 10^{-14}$  \\
 8  &  18.30  & 63.08  &  602.52  &  55574.74   \\  \hline
\end{tabular}
\end{table}

\begin{figure}[htb]
\centerline{
\includegraphics[width=15cm]{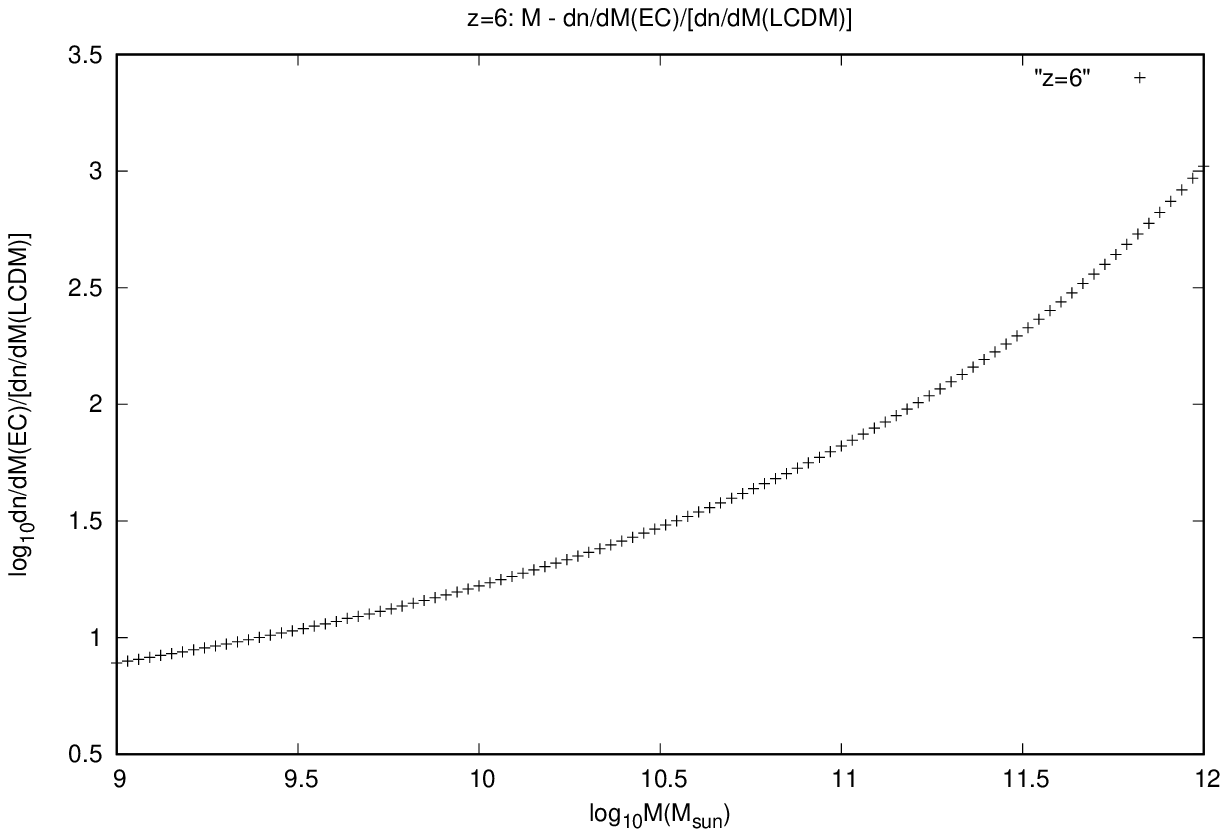}}
\caption{z=6:$log_{10}M(M_{\odot})$ vs. $log_{10}[\frac{dn}{dM}(EC)/\frac{dn}{dM}(\Lambda CDM)]$}       
\end{figure}

This phenomenon is observed few years ago (Fig.1 in \cite{Steinhardt}) and does not vanish with 
new observations \cite{newrefs}. A disparity between $\Lambda$CDM and EC models weighing
few orders of magnitude should not be considered as a surprise since the high-density 
regions follow probability rules in the form of the exponential fitting functions.
One can conclude that currrent measurements, albeit with limited statistics for higher redshifts,
support the EC model.

The first measurement of the cosmic 21-cm absorption signal from the neutral hydrogen
transition of EDGES \cite{EDGES} came as a surprise. Namely, the absorption signal
at redshift $z = 17.2$ appears to be much stronger than it is anticipated from
the standard brightness temperature contrast of the spin temperature against the CMB
\cite{Furlan,Loeb} for $\Lambda$CDM model:
 
\begin{eqnarray}
&&\delta T_{b} \approx 20mK (1-\frac{T_{cmb}(z)}{T_{s}(z)})\frac{x_{HI}(z)(1+\delta)}
{1+\frac{dv_{r}}{dr}/H(z)} \sqrt{\frac{1+z}{10}\frac{0.15}{\Omega_{m}h^{2}}}
\frac{\Omega_{b}h^{2}}{0.023}, \nonumber \\
&&\delta T_{b}(z \approx 17; EDGES) \approx -500 mK,\  
\delta T_{b}(z \approx 17; \Lambda CDM) \approx -220 mK. \nonumber 
\end{eqnarray}

One can reconcile the theory and measurements with cooling the intergalactic medium (IGM)
($ T_{k} \simeq T_{s} $) or adding to $T_{cmb}$ large radio background tenperature
$ T_{rad}$ at the EDGES signal redshift. Applying one of the standard recombination codes to 
the EC model, we estimate $ T_{k} $ at $z \simeq 17 $ \cite{recfast} to be only
$10\%$ cooler than $ T_{k} $ of $\Lambda$CDM. This is understandable from the
corresponding evolution equation for $ T_{k} $ (the second term with a Compton heating
has larger ionization fraction $\bar{x}_{i}$ in the numerator but also larger square root of 
the mass density parameter $H(z) \propto \sqrt{\Omega_{m}}$ in the denominator)\cite{recfast}:

\begin{eqnarray*}
\frac{d T_{k}}{d z} = \frac{2 T_{k}}{1+z} + \frac{8 \sigma_{T}u_{cmb}}{3m_{e}c(1+z)H(z)}
\frac{\bar{x}_{i}}{1+f_{He}+\bar{x}_{i}}(T_{k}-T_{cmb}).
\end{eqnarray*}

Therefore, the EC model has even smaller 21-cm absorption signal $\delta T_{b}(z \approx 17;EC)
\approx -100 mK$.

Feng and Holder \cite{Feng} proposed recently the inclusion of the additional radio background
that can resolve the problem with the EDGES signal, i.e. $ T_{radio} \simeq few\ T_{cmb} $.
We comment on 
three thorough analyses of the first galaxies \cite{Ewall,Mirocha,Qin} and their 
impact on the 21-cm signal. The first paper \cite{Ewall} investigates the first
black holes in mini-halos, their accretion rates and implications on radio, infrared,
X-ray and CMB backgrounds, as well their impact on possible heating of the intergalactic 
medium ($T_{k}$). The second paper \cite{Mirocha} deals with the early
atomic-cooling galaxies, modeling radiative processes and taking care of all
cosmological constraints. The third paper \cite{Qin} describes the first nonlinear
structures as molecularly cooled mini-halo galaxies. The extensive numerical analyses 
in these papers are performed within $\Lambda$CDM cosmology.

Although all three analyses in the $\Lambda$CDM model are based on large number of uncertain 
parameters, functions and procedures, one can emphasize the common conclusion:
it is possible to reproduce the EDGES absorption signal 
(to account for the excess radio background) but with very unusual
properties of galaxies at high-redshifts ($z \geq 17$), i.e.
$\simeq 10^{3}$ times more efficient star formation rates \cite{Mirocha} or 
$\simeq 10^{3}$ times greater X-ray production \cite{Qin}.

A key ingredient in analyses is the luminosity function
which is proportional to the number density of halos. If we compare 
number densities for mini-halos in EC and $\Lambda$CDM at redshift range
$z \epsilon [15,25]$ one can see in Table 3 the amplification of ${\cal O}(10^{2})-
{\cal O}(10^{4})$ in favour of the EC model.

\begin{table}
\caption{Mini-halo number density as a function of mass and redshift;
upper number=$\frac{d n}{d M}(EC;Mpc^{-3}M_{\odot}^{-1})$, lower number=$\frac{d n}{d M}(EC)$
/$\frac{d n}{d M}(\Lambda CDM)$.}
\vspace{3mm}
\hspace{10mm}
\begin{tabular}{| c || c | c | c | c |} \hline
        &             &                   &             &           \\
z $\backslash$ M($M_{\odot}$) & $10^{5}$ & $10^{6}$& $10^{7}$& $10^{8}$ \\  
\hline \hline
  & 0.359 &$3.45\cdot 10^{-3}$  & $4.32\cdot 10^{-5}$& $4.94\cdot 10^{-7}$  \\  
 15 & 9.35 & 17.15 & 60.02  & 214.68 \\  \hline
  & 0.413 &$3.98\cdot 10^{-3}$  & $4.98\cdot 10^{-5}$& $5.69\cdot 10^{-7}$  \\
 20 & 39.65 & 114.41 & 777.94  & 7662.22 \\  \hline
  & 0.46 &$4.45\cdot 10^{-3}$  & $5.57\cdot 10^{-5}$& $6.34\cdot 10^{-7}$  \\
 25  &  256.20  & 1315.19  &  20792.48  &  741775.22   \\  \hline
\end{tabular}
\end{table}

At the end of this section we point to new measurements and analysis \cite{Davies}
of ionizing photon mean free path at redshift $z=6$ resulting in "increased
demands on ionizing sources" in $\Lambda$CDM model. The EC model (see Table 2)
has larger number of ionizing sources than $\Lambda$CDM model.

\section{Conclusions}
With a great progress in precision and high-redshift measurements in astrophysics confronted 
with more powerful simulations and calculations, we are faced with severe problems
of the concordance $\Lambda$CDM cosmology. The EC cosmology with CDM and light neutrinos
represents a possible natural path to resolve problems of high-redshift Universe.
The effects of torsion through spin and angular momentum densities are backbones of the
new theoretical machinery. The violation of isotropy is already observed in the Planck
TT CMB spectrum \cite{Planck2}. There is a hint for a parity violation in polarization
data of Planck \cite{Japs,Korotky} with a preference to right-handedness \cite{Palle11}.
It could stimulate the theorists to undertake extensive numerical simulations 
with expanding and rotationg Universe to reveal the redshift dependence of torsion
from the first principles of the Einstein-Cartan cosmology.

\end{document}